# Proprioceptive Misestimation of Hand Speed

Caitlin Callaghan
Department of Mechanical and Aerospace Engineering
*University of California Irvine*
Irvine, CA United States
callaghc@uci.edu

David J. Reinkensmeyer
Department of Mechanical and Aerospace Engineering
Department of Anatomy and Neurobiology
*University of California Irvine*
Irvine, CA United States
dreinken@uci.edu

*Abstract*— The accuracy with which the human proprioceptive system estimates hand speed is not well understood. To investigate this, we designed an experiment using hobby-grade mechatronics parts and integrated it as a laboratory exercise in a large remote laboratory course. In a simple joint position reproduction task, participants (N = 191) grasped a servomotor-driven shaft with one hand as it followed a randomized trajectory composed of sinusoidal submovements. They simultaneously attempted to reproduce the movement by turning the shaft of a potentiometer with the other hand. Focusing on the first movement of the trajectory, we found that participants consistently overestimated the speed of the slowest rotations by ~45% and underestimated the speed of the fastest rotations also by ~30%. Speed estimation errors were near zero for trajectories with peak velocities ~63 deg/s. Participants' movements also overshot slow trajectories and undershot fast trajectories. We show that these trajectory errors can be explained by a model in which the proprioceptive system integrates velocity misestimates to infer position.

## I. Introduction

Proprioception, the body's sense of both position and movement in space, plays a major role in feedback control of movement [1], motor learning [2], and neurorehabilitation [3], [4], [5]. Proprioception of the hand is especially important for the control of fine movements, and loss of hand proprioception is a predictor of the ability to recover hand movement after stroke [3], [4]. Besides stroke, a wide variety of conditions and injuries, including Parkinson's Disease [6], Cerebral Palsy [7], and multiple sclerosis [8]. can impair proprioception. Characterizing how proprioception normally works is important for understanding how these conditions impair it.

Yet, despite the term "proprioception" referring to the sense of both position and movement, our current understanding of hand proprioception comes mainly from experiments that have focused on static positioning [9]. For example, in a recent experiment, Dandu et al examined the ability of individuals to estimate the positions of their fingertips when placed in different patterns against a tabletop. They found that static proprioception of the fingers is surprisingly inaccurate, with finger localization errors of multiple centimeters [10].

Sensing inaccuracies during movement are less well characterized, but initial studies suggest that proprioceptive error is larger for faster movements. For example, Goble and Brown found a velocity effect for memorized movements at the elbow [9], in which a manipulandum first moved the elbow through a triangular velocity profile with one of two maximum velocities, then participants immediately reproduced the movement. At higher speeds, participants were less accurate in replicating the desired maximum speed and their acceleration profiles were less symmetric than the template. Kerr and Worringham examined the ability to distinguish between two speeds during passive elbow extension [11], finding that the minimum difference to distinguish between speeds and the uncertainty of those discriminations increased with template velocity. In a mirroring paradigm, in which a robot passively guided one hand from one target to a random location while participants attempted to simultaneously copy the movement with their other hand, Tulimieri et al. found that fast movements led to larger position errors, but velocity error was highest at both fast and slow extremes [12].

These studies suggest the following hypothesis: the accuracy of proprioceptive perception of hand movement decreases at higher speeds. To evaluate this hypothesis, we designed a contralateral joint position reproduction task for the hand and wrist. Participants experienced a 5-second movement with their non-dominant hand, simultaneously reproducing the movement with their dominant hand. Deployment of the experiment as an educational experience in a large undergraduate engineering class allowed us to collect a large data set. We analyzed the relationship between template velocity and template reproduction error and found velocity sensing errors not only at high speeds, but also at low speeds. We then used the results of this analysis to develop a model of proprioceptive processing that accurately predicts trajectory errors based on the fundamental characteristics of velocity misestimation.

## II. Methods

### A. Experiment setting and participants

This experiment was incorporated into a junior level engineering laboratory course offered by the University of

C Callaghan is with the Mechanical and Aerosapce Engineering Department, University of California Irvine, Irvine, CA USA., callaghc@uci.edu

D Reinkensmeyer is with the with the Departments of Mechanical and Aerospace, and Anatomy and Neurobiology, University of California Irvine, Irvine, CA USA, dreinken@uci.edu

Research supported by NIH Grant 901HD062744 and NIDILRR Grant 90REGE0010

BioRobotics Lab, Department of Mechanical and Aerospace Engineering, University of California Irvine, Irvine CA 92617

California, Irvine in Spring 2021 during the pandemic when teaching was still remote. The educational objectives of the experiment were to: "Learn about feedback control by studying your own body's ability to implement and improve a proprioception feedback control loop" and "Learn common statistical techniques for analyzing data, including line and curve fitting, and testing for significant differences between data sets using the t-test." The experiment qualified for UC Irvine's Self-Determined Exemption as "Research conducted in educational setting involving normal educational practice" and participants provided informed consent, including whether their data could be used in a potential future scientific publication.

212 students were enrolled in the class. Students living in the United States were mailed a standardized kit of mechatronic parts to perform this and other laboratory exercises, but this was not possible for seven international students, who instead purchased locally available parts. Their data were omitted since their apparatus differed. Two students requested their data be omitted from the final dataset. 12 students submitted no or only partial data. The remaining 191 students' data were included in the analysis.

*B. Experimental protocol*

Participants assembled a test apparatus using the kit of mechatronics parts mailed to them (Fig. 1A), under the remote supervision of teaching assistants trained to supervise this experiment. A servomotor with position feedback (FEETECH Standard Servo FS5103B-FB) controlled by a microcontroller (Arduino UNO R3) haptically displayed template trajectories to the nondominant hand in 5-second runs. 80 possible templates were created, each the sum of four sine waves of random amplitude, frequency, and face shift (Fig. 1B). Template signals had at most 5 direction reversals across the 5-second duration (mean 3 ±1 reversals). 5 reversals were chosen because sequence retention weakens with more than 5 elements [13].

Participants attempted to reproduce the template with their dominant hand by holding a knob on a rotational potentiometer (Comimark WH148 Type B10K) and moving in the same pattern (Fig. 1B). They could do this by supinating/pronating their hand and/or twisting their fingers; we did not control the joints used. Participants were randomly assigned into two groups to practice primarily either simultaneous reproduction (SimRep) or memorized movement reproduction (MemRep). They were asked to perform 12 reproduction trials in their primary condition for four sequential days, along with 12 trials in their secondary condition on the first and last of the four days. Every time they performed the task, they experienced a new template drawn from the library of 80. In the SimRep condition, they moved simultaneously with the haptically displayed movement. In the MemRep condition, they moved following completion of the five second template trajectory.

Prior to data collection participants demonstrated that their apparatus was assembled according to provided diagrams and that the angular reference frames of servo and potentiometer matched via an Arduino program that used potentiometer position to control servo position. Participants confirmed that turning the potentiometer 90 degrees generated a 90-degree servo rotation in the same direction, and that the servomotor's position feedback function also reported 90 a degree rotation to the Arduino program. Participants performed the first set of 12 reproductions during the lab session, under the videoconferencing-based supervision of teaching assistants, to ensure they understood the task.

A visual interface (Matlab) guided participants through the experimental procedure. When the participant started the experiment, the servomotor moved to an initial angle, which participants were guided to match with cues to turn clockwise or counterclockwise. In the case that a participant finished a reproduction attempt within 3 deg of the next template's initial position, that template was swapped for another. This prevented a template movement initiating without the participant manually aligning potentiometer and servomotor position, which was the method for participants to signal their readiness for the next template to begin. As a result, not all templates were viewed by all participants, nor was there an equal number of reproduction attempts for each template. The mean (SD) number of students who attempted to reproduce each template was 95 (50);

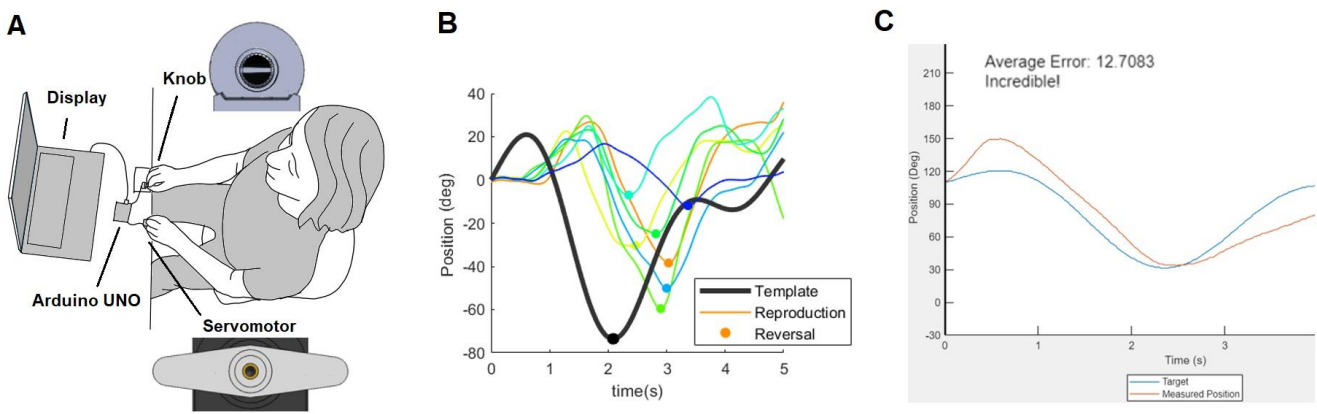

Figure. 1. Overview of experimental apparatus and data analysis. A: Participants assembled the testing apparatus from a kit containing an Arduino UNO microcontroller, a servomotor which displays the template movement, a potentiometer knob to sense hand position during the reproduction attempt, and their own computer, which displaying the experiment interface in Matlab. B: Example of template trajectory and reproduction attempts. Local peaks and troughs (direction reversals) are highlighted; the first reversal > 30 deg from the start position was used as the 'initial peak'. C: Example of feedback provided to participants after each trial. Further, after three runs, the previous three runs' results were displayed as the desired and actual positions, the mean position gap, and a rating was displayed to increase motive to do well. Three such plots were shown for 10 seconds before proceeding to the next template reproduction run.

templates experienced by fewer than 12 participants were not analyzed, leaving 60 templates for the below analysis.

Once potentiometer and servo were within 3 deg of the same position, a countdown was displayed for 3 s prior to template movement onset. The servomotor then displayed the 5 s template. After the run ended, there was a short countdown (3 s) before the servo moved to the start position of the next template. After three runs, a summary of the last three attempts was displayed for 10 seconds (Fig. 1C) as well as a rating message to increase motivation to track well.

The Matlab interface recorded data from the servo and potentiometer knob position at 60 Hz; position and time data were stored locally as files that were listed in the deliverables for the lab report associated with this experiment.

*C. Data analysis*

To estimate proprioceptive errors, we focused on the first submovement, defined as the portion of the trajectory between movement onset and the first local minimum or maximum > 30 deg from start position (Fig. 1B). The relative size of the position error was quantified as the ratio of the magnitude of the template's movement during this portion and that of the associated movement in the reproduction, the "Peak Magnitude Ratio" (PMR) (Fig. 2A). The relative size of the velocity error was quantified as the ratio of the maximum speeds of the template and reproduction during this initial movement, the "Peak Speed Ratio" (Fig. 2B). Relationships between template speed and position and velocity errors were quantified using Pearson's Correlation, after confirming template velocities were normally distributed (Kolmogorov-Smirnov test, p = 0.41).

*D. Developing a model of proprioceptive estimate of hand position*

We developed a model to generate simulated reproduction attempts and analyzed how well the simulated reproductions replicated human misestimation trends. This model was developed from the hypothesis that integration of velocity signals impacts position sense [14], and that this velocity signal would contain systematic errors equal to the ones we measured, in addition to noise [15], which we assumed was signal dependent [16], such that the variance of the velocity estimate was proportional to its magnitude.

From this basis, we developed a model of template reproduction as follows. For a given template, the model first converts the template position signal (sampled at 60 Hz) into a velocity signal (using the Matlab diff function). This is the signal that we assume the human proprioceptive system integrates to estimate position, except we also assume the signal has random and systematic errors. To simulate the random errors, white Gaussian noise was added to the velocity signal. Low-pass filtered Gaussian white noise is well-suited to nonlinear dynamic modeling of physiological systems [17]. To simulate systematic errors, the noise was biased dependent on whether the instantaneous template velocity was above or below 63 deg/s, the velocity accurate reproduction speed. The amplitude of the overall noise was also scaled by the magnitude of the template velocity (implementing signal-dependent noise). The resulting noisy velocity trajectory was then integrated into a position trajectory (using Matlab's cumtrapz function). The model is thus captured by the following equation:

$$\theta_{sim} = \int \dot{\theta}_{temp}(1 + A_1 * (N + A_2(C - |\dot{\theta}_{temp}|))) \, dt$$

where $\dot{\theta}_{temp}$ is velocity of the template, and $\theta_{sim}$ is the simulated reproduction trajectory. The parameters, which were hand-tuned ($A_1, A_2$) or found experimentally (C), are as follows:

N is Gaussian white noise (average = 0, variance = 1.0 )
$A_1$ = 2.0, the amplitude for the overall noise
C = 63 deg/s, the experimentally identified speed associated with accurate velocity reproduction
$A_2$ = 1/200 $\left(\frac{deg}{s}\right)^{-1}$, the gain for the bias to the noise. This gain shifted the fastest signals' noise (~112 deg/s) to average = -0.25, and the slowest (~23 deg) to average = 0.2.

Finally, we low pass filtered $\theta_{sim}$ to generate the simulated reproduction trajectory (fourth-order, low-pass Butterworth filter with a 1 Hz cutoff frequency) and time-shifted the signal the mean delay between template onset and the reproduction leaving the start position (reaction time), 0.27 sec.

100 such simulated reproductions were generated for each template. Position and velocity errors were quantified for each of the 100 reproductions, then averaged to obtain a predicted position and velocity error for that template.

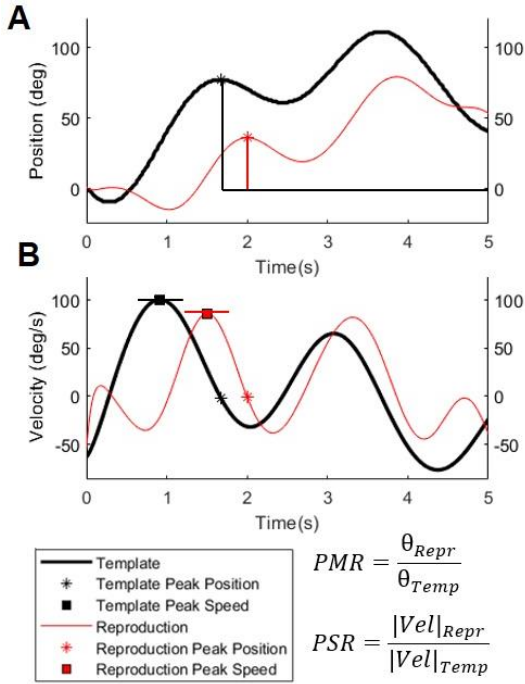

$$PMR = \frac{\theta_{Repr}}{\theta_{Temp}}$$

$$PSR = \frac{|Vel|_{Repr}}{|Vel|_{Temp}}$$

Figure. 2. Overview of error metrics. A: Error at the initial peak was quantified as the ratio of magnitude of template initial submovement, and magnitude of reproduction's initial movement (Peak Magnitude Ratio, or PMR). B. Velocity error was quantified as the ratio of replay and template's maximum speed for the movement during the first submovement, the 'Peak Speed Ratio'(PSR)

## III. Results

Participants (N = 191) used their dominant hand to try to reproduce rotational movements displayed to their non-dominant hand by a servomotor. The template movements followed randomized patterns composed by adding sinusoids.

### A. Reproduction errors of the first submovement in the simultaneous reproduction condition

Considering first the SimRep condition, the magnitude of the first reproduced movement strongly correlated with that of the template (Fig. 3A, R = 0.87, p < 0.001). However, participants tended to undershoot larger and overshoot smaller movements (Fig. 3A). Similarly, peak velocity of the reproduction strongly correlated with the peak velocity of the template (Fig. 3B, R = 0.75, p < 0.001), but participants overestimated the speed of slower rotations and underestimated the speed of faster ones.

The overshoot/undershoot pattern was related to the peak velocity of the first submovement of the template. When analyzed as a ratio, the magnitude of SimRep movements > 70 deg/s was underestimated by about 35% (PMR < 1.0), and the magnitude of movements with velocity < 70 deg/s was overestimated by ~45% (PMR > 1.0 Fig. 4A). The velocity of movements > 63 deg/s was underestimated by ~30% (PSR < 1.0), and the velocity of movements with velocity < 63 deg/s was overestimated by ~50% (PSR > 1.0 Fig. 4B). The mean accurate match speed across individuals was 63 ± 22 deg/s; 63 deg/s was taken as the accurate match speed for modeling purposes.

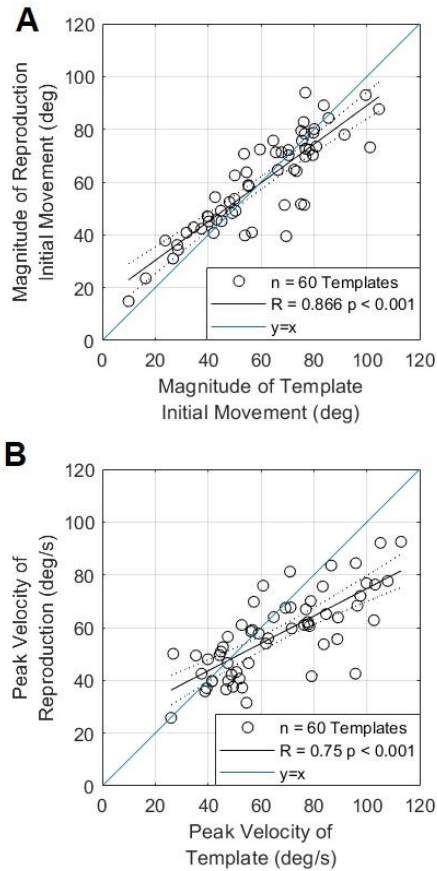

Figure. 3. Performance on the simultaneous reproduction task (SimRep), quantified by the accuracy of the first movement. A: The peak magnitude of the template and associated reproductions were highly correlated. Each point is the average across the individuals who practiced that template (average number of individuals = 95). B: The maximum velocity of the first submovement of a template was also strongly correlated with the maximum velocity of the reproductions associated with that template.

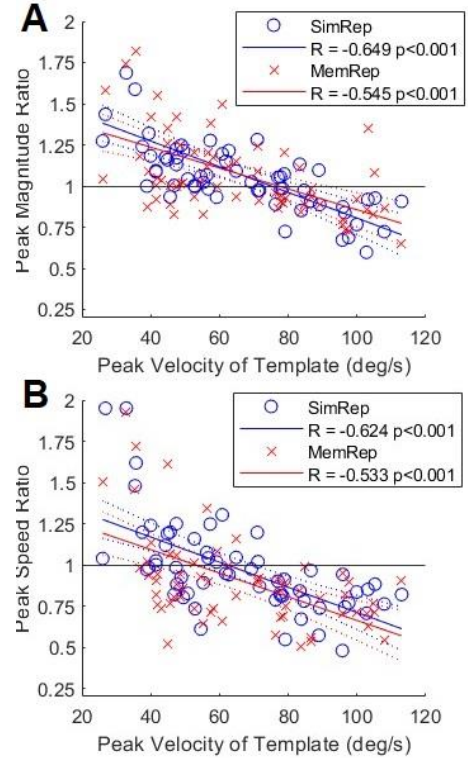

Figure 4. Performance on the simultaneous (SimRep) and memorized (MemRep) tasks quantified as ratios of reproduction to template magnitude (A) and speed (B) of the first submovement. A: Peak magnitude ratio. Linear fits were not significantly different for SimRep and MemRep. Position was accurately matched at ~73 deg/s. B: Peak speed ratio. Linear fits were also not significantly different, though accurate speed match was 63 deg/s for SimRep and 40 deg/s for MemRep. Linear regression fits compared with ANCOVA. p > 0.1 for all slopes and intercept comparisons.

We tested whether the magnitude of the template movement impacted this velocity-dependent, overshoot/undershoot pattern. To do this, we binned templates into four magnitude ranges (in terms of the first submovement) and performed correlation analysis of velocity and magnitude error within sets of templates with similar initial magnitudes. All bins showed a negative correlation between template velocity and overshoot (Fig. 5), these correlations had similar slopes (ANCOVA of regression lines, p > 0.05 for slope comparisons), indicating this relationship is similar regardless of movement magnitude.

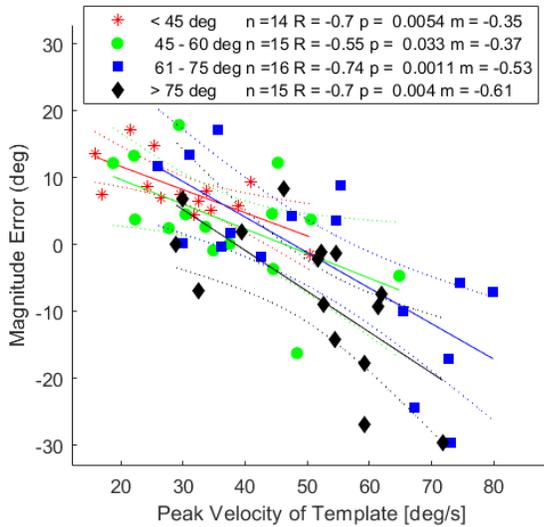

Figure 5. Analysis of whether first submovement amplitude affected the observed velocity-dependence of error. Negative correlation of template velocity and overshoot was present for templates of differing magnitudes. Templates were binned into groups with peak magnitudes equal to 0-45 deg, 46-60 deg, 61-75 deg, and > 75 deg.

### B. Velocity misestimation trends were similar for memorized and simultaneous reproduction

The analysis presented so far has focused on the simultaneous reproduction task (SimRep). Participants also tried to reproduce the template from memory, immediately after having haptically experienced it (MemRep). Magnitude and speed error trends for the first peak were similar during memorized reproduction (Fig. 4). Thus, the human sensory motor system apparently memorized the sensing misestimates, recreating them during the delayed reproduction in a similar way as when it had real-time access to haptic information about the template.

### C. Model

We created a model with the goal of replicating the observed position and velocity misestimation trends. We hypothesized that the proprioceptive system integrates velocity to estimate position, and therefore that errors in position misestimation would arise from systematic errors in velocity estimation. The model is fully described in the Methods, but, essentially, we modeled the experimental velocity estimation curve for the first submovement then used it to simulate position misestimation for both the first submovement, as well as the entire movement. The model successfully replicated the position (Fig. 6A) and velocity (Fig. 6B) error trends for the first submovement, as verified by ANCOVA comparison of regression fits ($p > 0.05$ for both slope and intercept comparisons). The mean error across the duration of the template was also similar to the mean error of participants (Fig. 6C, 6D), even though the model parameters had been selected based on the first submovement only.

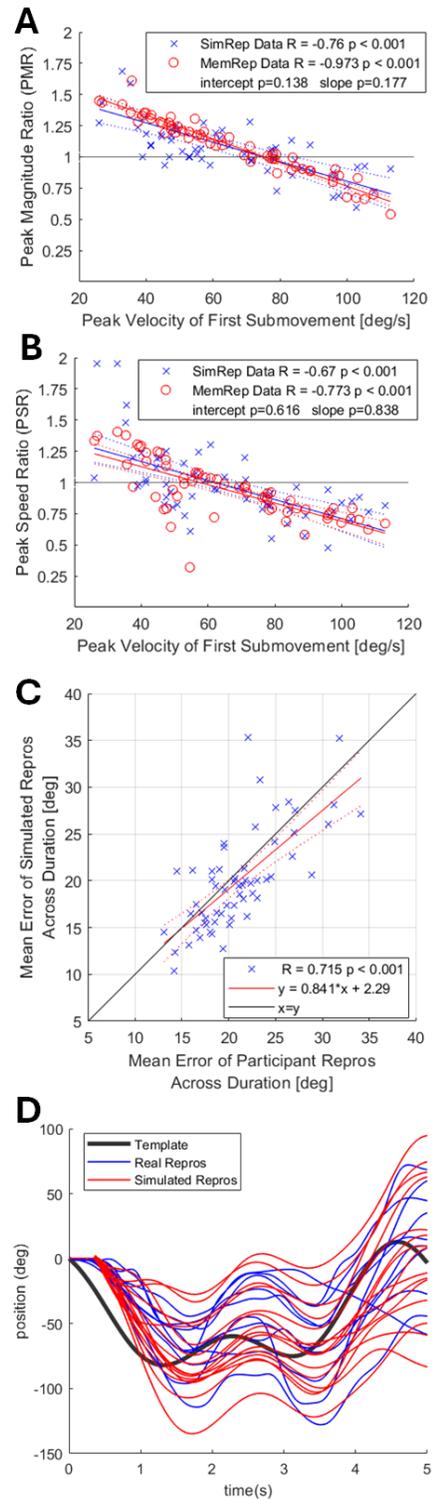

Figure 6. Simulated trajectories replicated the magnitude (A) and velocity misestimation (B) effects; fast movements were underestimated in both their magnitude and velocity, while slow movements were overestimated. C. The model well estimated mean error across the entire trajectory. Correlation between participants' mean error across the 5-second duration for a template and the error of simulated reproductions for the same template. D. A sample of real and simulated reproductions for one template.

## IV. DISCUSSION

The goal of this study was to evaluate the hypothesis that the accuracy of proprioceptive perception of hand movement is velocity dependent, with higher errors at higher speeds. We addressed this hypothesis through a bilateral hand position reproduction task: simultaneous and memorized reproduction of a set of templates with varying movement speeds and amplitudes. We found that both position and velocity estimations were systematically influenced by the velocity of the template being reproduced; fast speeds induced underestimations, and slow speeds induced overestimations. This effect was present in both simultaneous and memorized reproduction conditions. Simulated reproductions generated by a model incorporating these velocity-based errors replicated features of position errors. We discuss these results now, followed by limitations and future directions.

### A. Velocity-dependent proprioceptive accuracy

As reviewed in the Introduction, several previous studies have found that proprioceptive errors increased at higher speeds for various proprioceptive reproduction tasks [9], [11], [12]. In the current study, participants systematically underestimated both the magnitude and velocity of fast movements and overestimated them for slow movements. Thus, this study provides a more comprehensive description of the way movement speed affects proprioceptive estimates at both ends of the speed spectrum. The errors occur in such a way as if the human proprioceptive system has the wrong calibration "slope" or "gain" relating actual to perceived velocity. In general, this gain is too low, although the relationship is accurate at an intermediate speed.

Though not discussed in their analysis, Tulimieri et al's data (Figure 3) also suggests the existence of an accurate velocity reproduction speed (in their case it was ~ 0.15 m/s), with speeds slower than this being overestimated and faster speeds underestimated. As described earlier, their experiment involved a mirroring paradigm, in which a two degrees of freedom robot passively guided one hand from one target in the horizontal plane to a random location while participants attempted to simultaneously copy the movement with their other hand. Thus, proprioceptive misestimation of hand speed seems to apply both for multi-joint arm movements as well as the single DOF movement examined in the present study.

A difference between the current results and the Tulimieri results was that they found end point error was lowest at the slowest speeds, where, in the current study, end point error was lowest at an intermediate speed. The difference may arise from different "slowest" speeds or movement durations studied and should be examined in future work.

Although it appears that the proprioceptive system has too low of a "speed gain" when it converts motion to memory, a possible alternative explanation for the observed phenomenon, at least at higher speeds, is that the slower reproduction speed is a strategy to track the overall trajectory more accurately. If participants anticipated that the motor would soon reverse direction, then they may have predictively slowed down to "catch up" with the servomotor after it reversed direction. However, two observations argue against this possibility. First, the underestimation trend was present whether participants had access to real-time haptic feedback from the hand experiencing the template (SimRep) or not (MemRep). In the MemRep condition, there was no "catching up" to do, since individuals were replaying what they had already felt. Second, the underestimation trend was also present for point-to-point, horizontal plane hand movements in Tuilimieri et al. and these movements had a single velocity peak and no direction reversals. Thus, the errant velocity sensing gain seems inherent to the human sensory motor system, rather than an artifact of a gaming strategy.

### B. Velocity integration as a strategy for position estimation

The simple model of proprioceptive processing that we developed assumed that the proprioceptive system estimates positions during an imposed movement by integrating velocities. That is, the model integrates an estimated velocity trajectory to reproduce an experienced movement trajectory. Critically, the velocity estimate is noisy in two ways: 1) the noise is biased by the relative speed compared to an intrinsic speed which can be sensed accurately; and 2) the magnitude (and thus variance) of the noise scales with the velocity estimate. This model not only captured the experimentally-observed trends for the first submovement, but also predicted the mean error across the entire 5-second duration template.

The idea that the human sensory motor system estimates limb position by integrating an estimate of limb velocity has been suggested before, as in Clark et al.'s [14] development of velocity-only and velocity-displacement hybrid models for flexion and extension distinction in the fingers joints and ankle. That study found a velocity-only model adequate to predict sensitivity to flexion and extension for the fingers; here, we have also found a velocity-only model that can capture proprioceptive reproduction of hand position.

### C. Future directions

The reason why the human sensory motor system exhibits a single 'best' speed for movement sensation is unclear. An interesting direction for future research is to determine whether this accurate reproduction speed is consistent for different joints and proprioceptive tasks. For the task we studied, it would be interesting to compare this speed to movement speeds that individuals use during typical daily movement activities, including occupations, sports, and hobbies. This would test the hypothesis that velocity estimation is tuned to activity.

The reason why the human speed estimation gain is too low is unclear, and an interesting direction for future research. Perhaps it conveys an advantage in some way. One might argue that slowing down during the reproduction of fast movement occurs because of energy minimization considerations [18], but this doesn't explain why people speed up for slow movements. Clark et al. [14] predicted that velocity-only models would not be able to accurately capture slow movements, as there is a minimum sufficient velocity before movement can be accurately detected and assessed. Perhaps moving faster than the slowest template speeds arises from an effort to reach this sufficient velocity for accurate, proprioceptive sensing during movement. Future research on position and velocity estimation errors during very slow movements of short duration would test the hypothesis that there is a minimum speed the hand must experience before dynamic proprioception becomes accurate.

Finally, proprioceptive ability is known to be learnable, exhibiting several features similar to motor learning [2]. One recent study found that movement reproduction of passively-sensed trajectories improves with repeated experience [19]. Perhaps some of these proprioceptive learning phenomena can be attributed to the speed estimation gain becoming more accurate with training, a possibility that could be experimentally tested.